\documentclass[aip,apl,graphicx,floatfix,preprint]{revtex4-1}
\pdfoutput=1

\newcommand{\units}[1]{\ensuremath{\mathrm{#1}}}
\newcommand{\amount}[2]{\ensuremath{#1\:\units{#2}}}

\newcommand{\wo}{\ensuremath{\omega_{0}}}
\newcommand{\fo}{\ensuremath{f_{0}}}
\newcommand{\ck}{\ensuremath{C_{\kappa}}}
\newcommand{\tsub}[2]{\ensuremath{#1_{\mathrm{#2}}}}

\usepackage{graphicx}

\begin{document}





\title{Introduction of a DC Bias into a High-$Q$ Superconducting Microwave Cavity} 








\author{Fei Chen}

\email[]{Fei.Chen@dartmouth.edu}




\affiliation{Department of Physics and Astronomy, Dartmouth College, Hanover, New Hampshire 03755}

\author{A. J. Sirois }





\affiliation{University of Colorado, 2000 Colorado Ave,
Boulder, Colorado 80309, USA}

\author{R. W. Simmonds}





\affiliation{National Institute of Standards and Technology, Boulder, Colorado 80305}

\author{A. J. Rimberg}

\email[]{Alexander.J.Rimberg@dartmouth.edu}




\affiliation{Department of Physics and Astronomy, Dartmouth College, Hanover, New Hampshire 03755}

\date{\today}

\begin{abstract}

We report a technique for applying a dc voltage or current bias to the center conductor of a high-quality factor superconducting microwave cavity without significantly disturbing selected cavity modes. This is accomplished by incorporating dc bias lines into the cavity at specific locations. The measured S-matrix parameters of the system are in good agreement with theoretical predictions and simulations. We find that at \amount{4}{K} the quality factor of the cavity  degrades by less than $1\%$ under the application of a dc bias.

\end{abstract}


\maketitle 

The circuit quantum electrodynamics (QED) architecture, consisting of a coplanar-waveguide-based superconducting microwave cavity, has been demonstrated to realize cavity QED physics on a chip.\cite{Wallraff:2004} This architecture has been successfully employed in quantum information processing,\cite{Wallraff:2004,Majer:2007,Sillanpaa:2007,Dicarlo:2009} and by integration with a nanomechanical resonator has allowed quantum measurement\cite{Regal:2008,Hertzberg:2009} and cooling of nanomechanical motion.\cite{Rocheleau:2010} Although the existing circuit QED architecture is quite versatile, the ability to apply a dc bias to its center conductor would make if more so.  For instance, a nanomechanical resonator strongly coupled to a qubit can also be embedded within such a  cavity to allow study of decoherence of macroscopic objects.\cite{Blencowe:2008}  In this letter, we demonstrate a technique for applying a dc voltage or current bias to the center conductor of a coplanar waveguide (CPW) cavity without significantly disturbing a particular cavity mode or degrading its quality factor Q at high frequencies. 

\begin{figure}[htbp]

\includegraphics[width=3.15in]{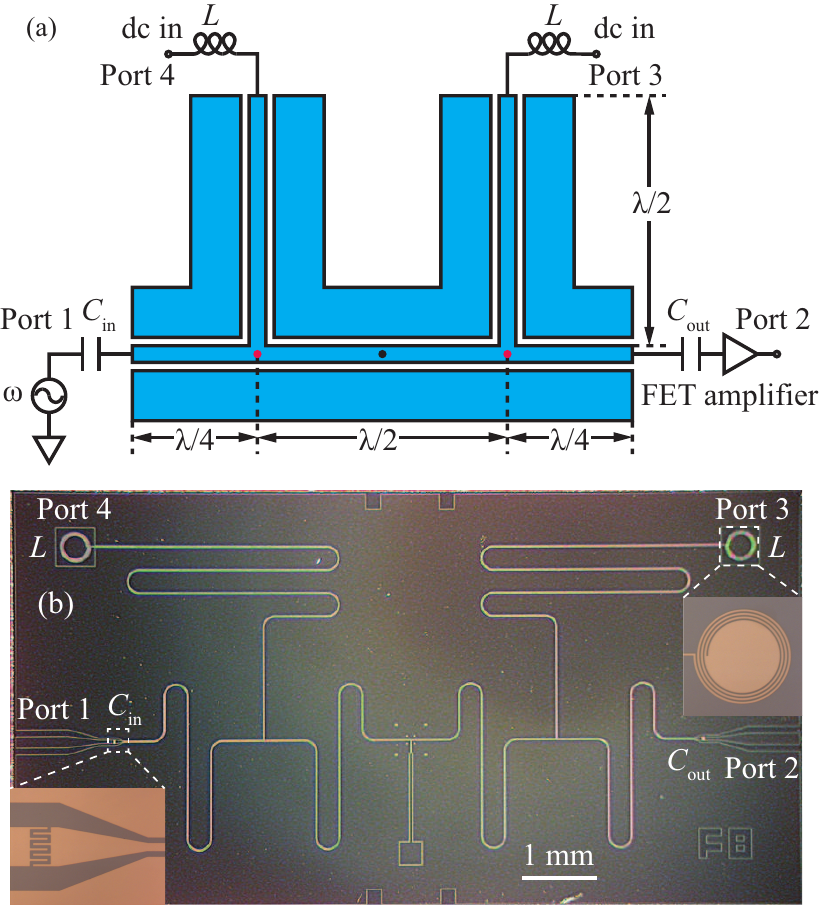}%

\caption{\label{fig1}(a) Schematic diagram for introduction of a dc bias into a high-Q microwave cavity of length $\lambda$. The red dots are the low impedance points and the black dot is a high impedance point. A Cooper pair box or other structure can be located at the high impedance point. (b) Optical micrograph of a \amount{4.8}{GHz} superconducting niobium coplanar waveguide resonator with dc feed lines. The insets show an 8-finger capacitor and a 3-turn spiral inductor.}

\end{figure}

The dc biasing scheme is shown schematically in Fig. \ref{fig1}(a). We choose the main  transmission line length $l$ to be a full wavelength ($l=\lambda$) at the resonant frequency $f=\fo$. To allow application of a dc bias, two sections of $\lambda/2$-long transmission lines are added to the main full-wave resonator at points a distance $\lambda/4$ from either end (marked with the red dots), forming two ``T'' junctions. Each of the $\lambda/2$ shunt transmission lines is connected via an inductor $L$ to a dc voltage or current source.

To understand the idea behind this arrangement, we first assume the main $l=\lambda$ transmission line is lossless and terminated in an open circuit. The current at the end of the line is then zero and the voltage is a maximum, making this a high impedance point. The $\lambda/4$ point is then a voltage node (current antinode) and a low impedance point.  Also, as is well known from standard transmission line theory,\cite{Pozar:2005} a $\lambda/2$ length of transmission line has an input impedance equal to that of its terminating impedance. The  inductively terminated $\lambda/2$ dc bias lines therefore present a high impedance $i\wo L$ where $\wo = 2\pi\fo$ to the main transmission line at the resonant frequency \fo.   A microwave photon approaching the $\lambda/4$ point from the center of the cavity will therefore see a short (the $\lambda/4$ line to the end of the cavity) in parallel with a high impedance (the inductively terminated $\lambda/2$ bias line). To a first order approximation then, the bias lines will have no effect on the full-wave cavity resonance. 

In reality, the transmission line is not completely lossless, and the main line is terminated with small capacitors leading to \amount{50}{\Omega} transmission line rather than with an open circuit. The impedance looking from a low impedance point toward either end of the line is still small but not identically zero.  Reasonable values of inductance $L$ must ensure the dc feed lines present a sufficiently large impedance to the main line for the approximate picture described above to be valid.

Fig. \ref{fig1}(b) shows a Nb coplanar waveguide resonator with dc feed lines. The CPW central conductor width is \amount{10}{\mu m} and its separation from the ground planes is \amount{5.5}{\mu m}, giving a wave impedance of approximately \amount{50}{\Omega}.   The resonator is fabricated on a $10.5\times\amount{5.5}{mm^{2}}$ high resistivity silicon chip using sputtering, photolithography and dry etching, with an overall length $l=\amount{24}{mm}$, giving $\fo\approx\amount{4.8}{GHz}$. The resonator is coupled via identical capacitors $\tsub{C}{in} = \tsub{C}{out}=\ck$ at each end to input and output microwave lines. Two coupling capacitors with different geometries are used to test the cavity design, namely 4-finger ($\ck\approx\amount{4.4}{fF}$) and 8-finger ($\ck\approx\amount{11.1}{fF}$) capacitors. A 3-turn spiral inductor (linewidth \amount{5}{\mu m}, line spacing \amount{10}{\mu m}, $L\approx\amount{5}{nH}$) is used to terminate the $\lambda/2$ shunt transmission lines.

\begin{figure}[thbp]

\includegraphics[width=3.15in]{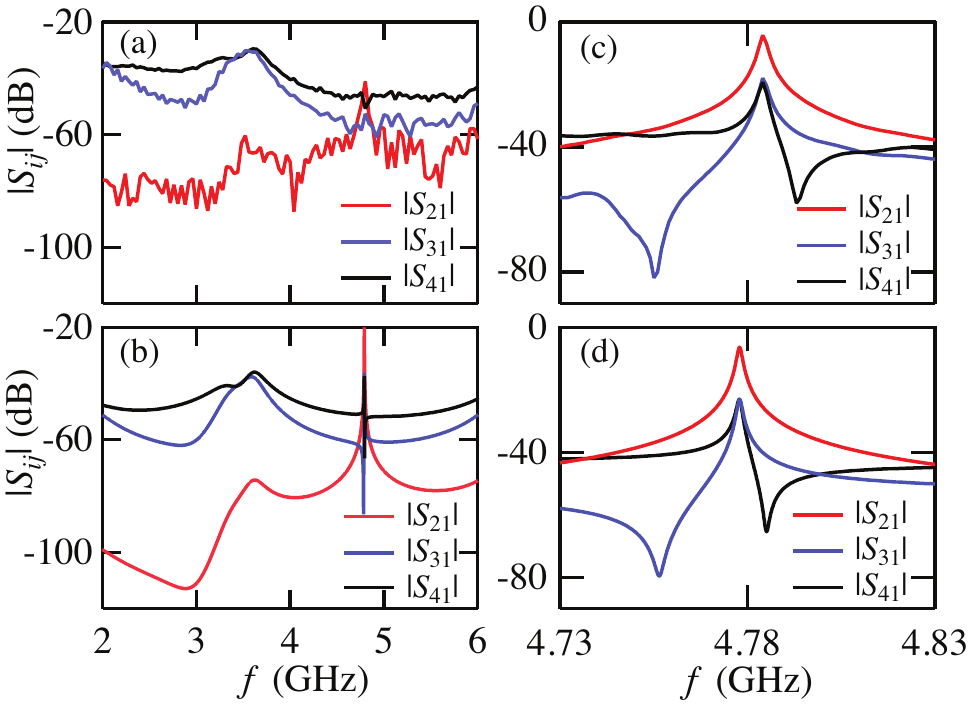}%

\caption{\label{fig2}(a) Measured and (b) simulated transmission spectra $S_{21}$, $S_{31}$ and $S_{41}$ of Resonator A from 2 to \amount{6}{GHz}. (c) Measured and (d) simulated transmission spectra $S_{21}$, $S_{31}$ and $S_{41}$ of Resonator B in the immediate vicinity of the full wave resonance.}

\end{figure}

Using a signal generator and a spectrum analyzer, transmission measurements of resonators A and B (4- and 8-finger coupling capacitors, respectively) were performed at 4.2 K inside an Amumetal 4K magnetic shield.  A room temperature FET amplifier (gain \amount{45}{dB}) was used to amplify the signal prior to measurement by the spectrum analyzer.  Losses in the cables leading to and from the sample were carefully measured and accounted for in the transmission measurements.  For all measurements unused ports were terminated by a \amount{50}{\Omega} impedance.

To characterize the microwave performance of the system, we measure transmitted power from the input capacitor (port 1) through the output capacitor (port 2) as well as the power through each spiral inductor (port 3 and 4). Fig. \ref{fig2}(a) shows measured S-matrix parameters $S_{21}$, $S_{31}$ and $S_{41}$ of resonator A from 2 GHz to 6 GHz.  Due to the presence of the dc feed lines, the fundamental $\lambda/2$ resonance  at around 2.4 GHz is strongly damped and is not visible in $S_{21}$.  This is to be expected, since for the fundamental resonance the bias lines are not located at a low impedance point.   In contrast, the first harmonic at \amount{4.8}{GHz} corresponding to a full wave excitation still exists as predicted by the simple impedance-based argument above.  Furthermore, additional broad resonances are visible at around \amount{3.6}{GHz} in $S_{21}$, $S_{31}$ and $S_{41}$, as well as more weakly in $S_{41}$ at about \amount{3.2}{GHz}. 

To investigate the origin of these additional resonances, we simulated the transmission spectra of the resonators using Microwave Office, as shown in Fig. \ref{fig2}(b).  The intrinsic $Q=6100$ of the resonators was chosen to agree with that of a separately measured undercoupled resonator, and is likely limited by losses in the Nb films at the relatively high temperature (\amount{4.2}{K}) of the measurements. The dielectric constant ($\varepsilon=13.1$) of the substrate was chosen to give the best agreement between the measured resonant frequencies of both resonators A and B.  The simulations show good agreement with experimental data over a broad frequency range as shown for resonator A in Fig.~\ref{fig2}(b).  The full wave resonance of the main line at \amount{4.8}{GHz} in Fig. \ref{fig2}(a) appears smaller than theory (as in Fig.~\ref{fig2}(b)) due to the relatively coarse sampling used in the measurement.

The broad resonances at 3.2 and 3.6 GHz are also clearly reproduced in the simulations.  Comparison between the simulations and data allows us to identify these resonances as arising from the fundamental resonance of the $3\lambda/4$ length of CPW between port 1 and port 4, and from the first harmonic resonance of the $5\lambda/4$ length of CPW between port 1 and port 3, respectively (see Fig.~\ref{fig1}(a)).   The resonant frequency of the $5\lambda/4$ section of CPW is pulled downward significantly by the reactance of the $\lambda/4$ section of transmission line\cite{Pozar:2005} from the second bias line to the output at port 2.  

To more carefully study the characteristics of the resonators near the full wave resonance we measured and simulated the same $S$ parameters over a much narrower frequency range immediately around \fo, as shown in Fig.~\ref{fig2}(c) and (d) for resonator B; agreement between the measurements and simulations is excellent.  The resonance in $S_{21}$ for the main line has a narrow Lorentzian lineshape (see below). Transmission through the dc bias lines as shown in Fig. \ref{fig2}(c) for $S_{31}$ and $S_{41}$ displays more complex behavior.  These resonances are well described by the Fano lineshape,\cite{Fano:1961} which generically results from mixing between a discrete state and a continuum. In our case, the continuum arises from the tails of the broad resonances associated with the dc bias lines, while the discrete state is the narrow full wave resonance of the main line.   When a photon is transmitted to one of the dc bias lines it may either couple directly to the bias line, or first couple to the main resonance and then to the bias line.  Interference between these two pathways is destructive on one side of the main resonance and constructive on the other, leading to the characteristic Fano lineshape.

This analysis suggests an alternative interpretation of the operation of the dc-biased cavity.  The dc bias lines are in fact  additional microwave resonators whose resonant frequencies are far from that of the main line.  The bias line resonators necessarily have a low $Q$ due to their strong  dc coupling to the measurement circuitry; as a result, their resonances still have significant weight at the frequency \fo\ of the main resonance.  Mixing with the sharp main line resonance occurs at \fo, allowing some fraction of the photons in the main line to escape via the bias lines.  Nonetheless, the mixing is relatively weak: as can be seen in Fig.~\ref{fig2}(c) and (d), the peak values of $S_{31}$ and $S_{41}$ at \fo\ are about \amount{13}{dB} less than that of $S_{21}$, indicating that less than $2.5\%$ of the input power escapes through the dc bias lines.   Coupling to the bias lines can be further weakened by using a larger terminating inductance $L$, which increases the bias line $Q$ and narrows their resonances.  For instance, using $L=\amount{20}{nH}$ increases the bias line isolation to roughly \amount{20}{dB}.

\begin{figure}[thbp]

\includegraphics[width=3.15in]{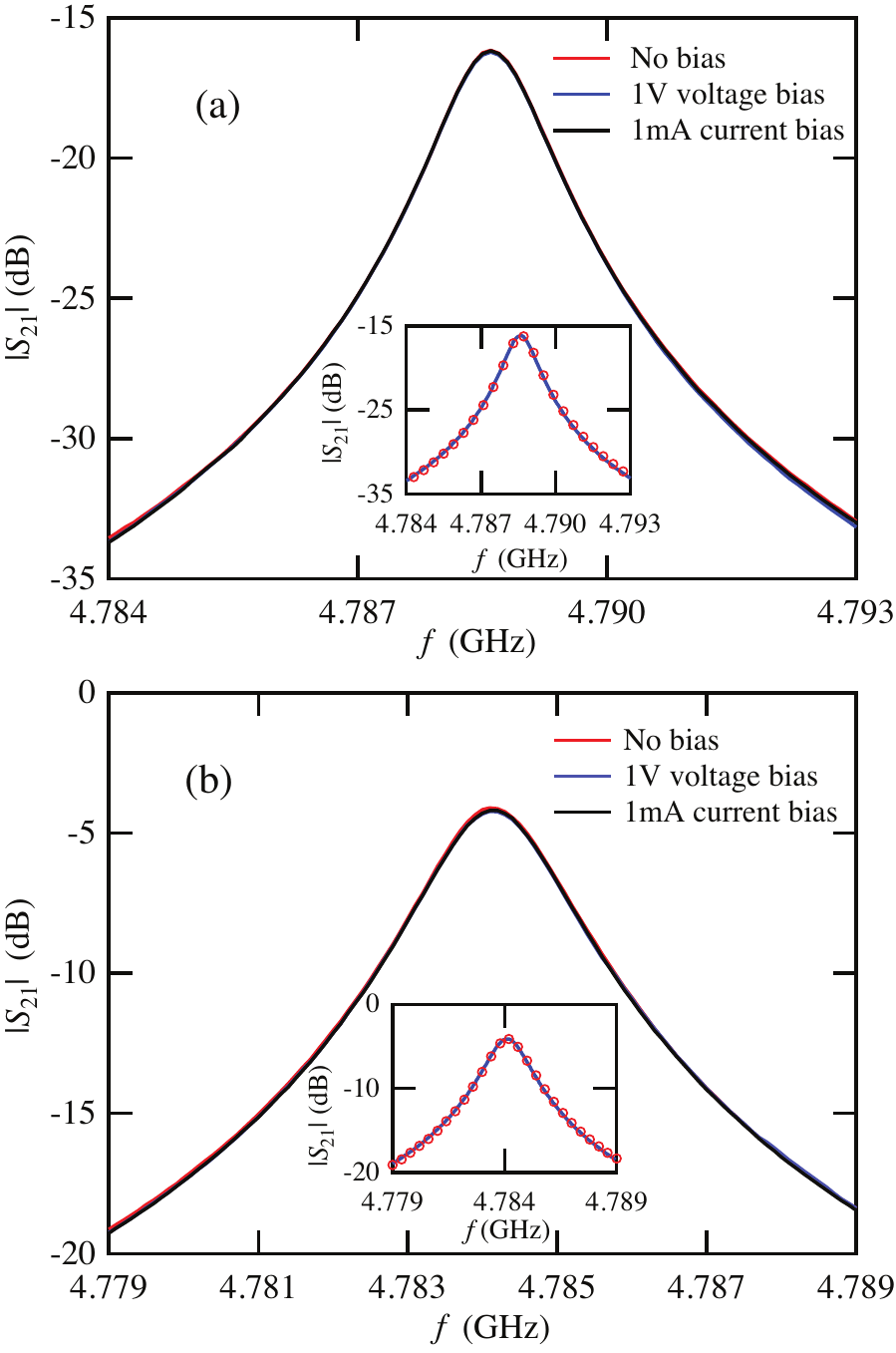}%
\caption{\label{fig3}(a) Measured transmission spectra $S_{21}$ of (a) resonator A  and (b) resonator B under different bias conditions. Insets: Lorentzian curve fits (blue curves) to the data (red circles) when no dc bias is applied. To show the curve fits clearly, the density of data points displayed in the insets has been reduced by a factor of 4.}%

\end{figure}

We now examine the effect of a dc bias on the main line full wave resonance.  We determine the quality factors of the resonances by fitting the $S_{21}$ spectra to a Lorentzian lineshape, as shown in the insets of Fig. \ref{fig3}(a) and (b). The data are clearly well described by Lorentzians with center frequencies of around \amount{4.8}{GHz} and quality factors as determined from the curve fits of $Q_A=3750$ and $Q_B=2500$. These values agree reasonably well with the total $Q$ expected from the intrinsic $Q$ and the loading due to the coupling capacitors.\cite{Goppl:2008} The addition of the dc feedline structure therefore has no significant effect on the $Q$ of the first harmonic mode at the temperature (\amount{4}{K}) of our measurements.

Finally, we apply a \amount{1}{V} dc voltage bias or \amount{1}{mA} dc current bias to the center conductor of the resonators via the bias lines. The resulting transmission $S_{21}$ spectra are nearly unperturbed by the dc bias, as shown in Fig. \ref{fig3}(a) and (b). The quality factors of the resonators as determined by fits to a Lorentzian lineshape degrade by less than $1\%$ under application of dc bias. We conclude that application of a dc bias does not significantly disturb the first harmonic cavity mode, as expected for our design.

The ability to introduce a dc bias into a high-$Q$ microwave cavity is an important addition to the circuit QED architecture. With dc access to the cavity, new highly non-linear devices can be developed by embedding current- or voltage-biased Josephson structures such as SQUIDs or single electron transistors\cite{Xue:2009} in the cavity. The dc-biased cavity may also serve as a general platform allowing very strong coupling between the cavity and a variety of quantum systems such as qubits, nanowires and nanomechanical resonators.


\begin{acknowledgments}

We thank M. P. Blencowe and M. Dykman for helpful conversations.  This work was supported by the NSF under Grant No.\ DMR-0804488 .

\end{acknowledgments}



%
\end{document}